\documentclass[12pt]{article}
\usepackage{latexsym,graphics,graphicx,amssymb,amsmath, amstext, amsfonts, amsthm, pifont, color,subfig}

\newcommand{\nn}{\nonumber}
\newcommand{\bc}{\begin{center}}
\newcommand{\ec}{\end{center}}
\newcommand{\bfl}{\begin{flushleft}}
\newcommand{\efl}{\end{flushleft}}

\newcommand{\beqa}{\begin{eqnarray}}
\newcommand{\eeqa}{\end{eqnarray}}
\newcommand{\beqan}{\begin{eqnarray*}}
\newcommand{\eeqan}{\end{eqnarray*}}
\newcommand{\beq}{\begin{equation}}
\newcommand{\eeq}{\end{equation}}
\newcommand{\beit}{\begin{itemize}}
\newcommand{\eeit}{\end{itemize}}

\newcommand{\lbr}{\left \{ }
\newcommand{\rbr}{\right \} }

\newcommand{\lp}{\left (}
\newcommand{\rp}{\right )}
\newcommand{\df}{\stackrel{{\rm def}}{=}}
\newcommand{\eqo}{\stackrel{.}{=}}
\newcommand{\lto}{\stackrel{.}{<}}
\newcommand{\gto}{\stackrel{.}{>}}
\newcommand{\leqo}{\stackrel{.}{\leq}}
\newcommand{\geqo}{\stackrel{.}{\geq}}
\newcommand{\real}{{\mathbb{R}}}

\newcommand{\prob}{{\mathbb{P}}}

\newcommand{\C}{\mathcal}

\newtheorem{lemma}{Lemma}

\newcommand{\A}{\mathcal{A}}

\newcommand{\SNR}{{\rm SNR}}
\newcommand{\INR}{{\rm INR}}

\newtheorem{thm}{Theorem}

\makeatletter

\author{Adnan Raja  and Pramod Viswanath}
\title{Diversity-Multiplexing Tradeoff of the Two-User Interference Channel}
\date{\today}

\begin{document}
\maketitle

\begin{abstract}
Diversity-Multiplexing tradeoff (DMT) is a coarse high SNR
approximation of the fundamental tradeoff between data rate and
reliability in a slow fading channel. In this paper, we characterize
the fundamental DMT of the two-user single antenna Gaussian
interference channel. We show that the class of multilevel
superposition coding schemes universally achieves (for all fading statistics)
 the DMT for the two-user interference channel. For the
special case of symmetric DMT, when the two users have identical
rate and diversity gain requirements, we characterize the DMT
achieved by the Han-Kobayashi scheme, which corresponds to two level
superposition coding.
\end{abstract}

\section{Introduction}\label{sec:intro}

\begin{figure}[h]
\begin{center}
\scalebox{0.65}{\input{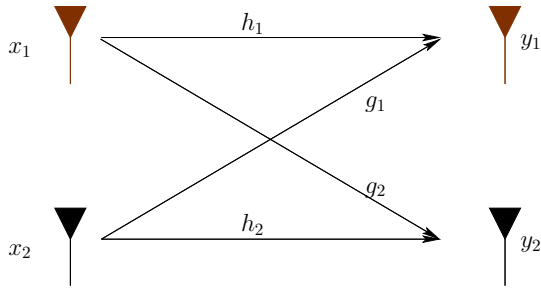}}
\end{center}
\caption{The slow fading two-user Gaussian interference channel. }
\label{fig-GICmodel}
\end{figure}

Consider the communication scenario depicted in
Figure~\ref{fig-GICmodel}. Two transmitter-receiver pairs
communicate reliably in the face of mutual
 interference with each other.
There is a single antenna at each transmitter and receiver. The
discrete time complex baseband model is: \beqa y_1[m] & = &
h_{1}x_1[m] + g_{1}x_2[m] + z_1[m], \label{eq:IFCmodel1} \\
y_2[m]&=&h_{2}x_2[m]+g_{2}x_1[m]+z_2[m].  \label{eq:IFCmodel2} \eeqa
Here $m$ is the time index, $y_k$ is the signal at receiver $k$
while $x_k$ is the signal sent out by the transmitter $k$ (with
$k=1,2$). The noise sequences $\lbr z_1[m], z_2[m]\rbr_m$ are
i.i.d.~complex Gaussian with zero mean and variance $N_{0}$. The
transmitters are subject to average power constraints: \beq
\sum_{m=1}^N |x_k[m]|^2 \leq NP_k, \quad k=1,2, \quad \forall N \geq
1. \eeq

The complex parameters $\lp h_{1}, g_{1}, h_{2}, g_{2} \rp$ model
the channel gain coefficients. Note that $\lp h_{1},g_{1} \rp$ are
the channel gain coefficients corresponding to receiver 1 and $\lp
h_{2},g_{2} \rp$ are the channel gain coefficients corresponding to
receiver 2. A rate pair $(R_{1},R_{2})$ is said to be achievable if
transmitter $i$ can send message to receiver $i$ $(i=1,2)$ at rate
$R_{i}$ with arbitrarily large reliability. The set of all
achievable rate pairs is the capacity region of the channel. In
\cite{ETW08}, it was shown that there are schemes that achieve
within 1 bit per unit time of the capacity of this channel. However,
these schemes assume that the channel gain coefficients are known
exactly at the transmitters.

The focus of this paper is the  \emph{slow fading} scenario, i.e., the
channel gains  though random, remain constant over the communication
time scale. Correspondingly, we will suppose that the receivers know
the channel coefficient gains. In the absence of any feedback link,
the transmitters do not know the channel gains; they are aware
of the channel statistics, however.  This classical setting has received much
attention in the context of point to point channels (a classic work
is \cite{ZT03}, but is also text book material -- Chapter~9 of \cite{TV05})
and multiple access channels \cite{TVZ04}.

Our focus is on the fundamental tradeoff between data rates (for
each transmitter) and reliabilities (for each receiver). In the
context of slow fading, the fundamental quantity of interest is the
{\em outage capacity region}: for a given unreliability of
communication, this is the pair of rates at which simultaneous
reliable communication is possible. Outage capacities are hard to
characterize exactly for even the point-to-point channel (formulas
have been derived for single antenna channels with specific channel
statistics, such as Rayleigh fading). A coarser characterization of
this tradeoff is captured by the diversity-multiplexing tradeoff (DMT),
first introduced for the point-to-point channel in \cite{ZT03}. The
diversity-multiplexing tradeoff is essentially a high SNR
approximation, of both the data rate and reliability. The
multiplexing gain is defined as \beq r \df
\lim_{\SNR\rightarrow\infty}\; \frac{R}{\log\SNR}. \eeq The
diversity gain is defined as \beq d \df
\lim_{\SNR\rightarrow\infty}\; - \frac{\log \prob_{out}}{\log\SNR}.
\eeq Stated imprecisely,
\begin{align}
R &\approx r \log\SNR, \label{eq:ordRNotation}\\
\prob_{out} &\approx \SNR^{-d}. \label{eq:orddNotation}
\end{align}
The tradeoff between $d$ and $r$ is the diversity-multiplexing
tradeoff.

In the context of the slow fading two-user interference channel, we
are interested in finding the largest set of rates of communication
$R_{1}$ and $R_{2}$ for the two users while tolerating an outage of
$P_{1,out}$ and $P_{2,out}$ at the two receivers respectively. In
the context of DMT, we want to find the fundamental tradeoff between
the multiplexing gains $r_1,r_2$ and the diversities $d_1,d_2$ which
are defined along the lines of \eqref{eq:ordRNotation} and
\eqref{eq:orddNotation}.

Recent work by Akuiyibo and L\'{e}v\^{e}que \cite{AL08} has
characterized some outer bounds for the DMT of the interference
channel. Our results show that these bounds are weak in most cases.
In \cite{AB09} Ak\c{c}aba and B\"{o}lcskei have given an achievable
DMT region for the case of very strong interference. Their analysis
deals mostly with finite block length schemes.

In this paper, we will allow schemes with arbitrarily large block
lengths. Our main result is the characterization of the fundamental
DMT for the scalar two-user interference channel. We do this by
defining suitable channel outage events for both receivers.  When in
outage, the receivers would very likely be in error. However, when
the channel is not in outage, the compound channel result of
\cite{RPV08} ensures that a multilevel superposition coding scheme
with opportunistic decoding of interference achieves the required
rate. Hence, we show that this scheme achieves the DMT universally
(for all channel statistics).

The Han-Kobayashi scheme is a special case (when the
number of levels are restricted to two) of the more general multi-level
superposition coding scheme introduced in \cite{RPV08}. We characterize the
performance of the Han-Kobayashi scheme in the context of the
special case of symmetric DMT, where the two users have identical
rate and diversity requirements. We compare this  symmetric DMT
achievable scheme by this scheme, to an outer-bound obtained by
considering the worst case channel outer bound of the compound
interference channel. While the Han-kobayashi scheme is DMT-optimal in some
instances, we also see that it is not DMT-optimal in many other cases.

The rest of this paper is organized as follows. In Section
\ref{sec:model} we quickly describe the system model and
assumptions. We characterize the fundamental DMT of the two-user
interference channel by relating it to the compound interference
channel problem in Section \ref{sec:main}. In Section
\ref{sec:symCase} we characterize the performance of the
Han-Kobayashi scheme for the case of symmetric DMT, where the rate
and diversity requirements for the two users are identical. In
Section \ref{sec:spCases} we consider two important special cases -
the Z-interference channel and the interference channel with only
the cross links fading - for which the performance of the
Han-Kobayashi scheme {\em is} DMT optimal.

\noindent{\em Notation:} We will use $(a\vee b)$ to denote the
maximum of $a$ and $b$, and $(a\wedge b)$ to denote the minimum of
$a$ and $b$. $(x)^+$ will be used to denote $(x\vee 0)$. $f(x) \eqo
g(x)$ will be used to denote \beq \lim_{x\rightarrow\infty}
\frac{\log f(x)}{\log g(x)} = 1. \eeq The symbols $\lto,\gto,\leqo$
and $\geqo$ will be used in the same spirit. $\prob(.)$ will be used to denote the probability of an event.

\section{Model} \label{sec:model}

The two-user interference channel model described in
\eqref{eq:IFCmodel1} and \eqref{eq:IFCmodel2} can be written in the
following equivalent form. \beqa \tilde{y}_1[m] & = &
h_{1} \sqrt{\SNR_1} \tilde{x}_1[m] + g_{1} \sqrt{\INR_1} \tilde{x}_2[m] + \tilde{z}_1[m],\\
\tilde{y}_2[m]&=&h_{2} \sqrt{\SNR_2} \tilde{x}_2[m]+g_{2}
\sqrt{\INR_2} \tilde{x}_1[m]+\tilde{z}_2[m]. \eeqa Note that
$\SNR_k$ and $\INR_k$ are the average signal-to-noise and
interference-to-noise ratios at receiver $k$ respectively ($k=1,2$). The
normalized i.i.d.~noise sequences $\lbr \tilde{z}_1[m],
\tilde{z}_2[m]\rbr_m$ have unit variance and the normalized transmit
sequences $\lbr \tilde{x}_1[m], \tilde{x}_2[m]\rbr_m$ are subject to a unit
average power constraint.

We will assume that the channel gain coefficients, $\lbr h_{1},
g_{1}, h_{2}, g_{2} \rbr$, are also normalized and are zero mean and
unit variance random variables. We will not suppose any particular
statistical distribution for the channel gains. We will only assume
that the channel gains exhibit an exponential tail, i.e., for
$X\in\lbr h_{1}, g_{1}, h_{2}, g_{2} \rbr$ there exists a $\gamma>0$
such that \beq \lim_{x\rightarrow\infty} \frac{\mathbb{P} \lp
|X|^{2}\geq x\rp}{e^{-\gamma x}} \leq 1. \label{eq:expTail} \eeq The
DMT will be characterized by the near-zero behavior of the
distribution functions, which is characterized by $\kappa_{X}$ for
$X\in\lbr h_{1}, g_{1}, h_{2}, g_{2} \rbr$ and is given by, \beq
\kappa_{X}\df \lim_{\epsilon\rightarrow 0+} \frac{\log \mathbb{P}\lp
{|X|^{2}} < \epsilon \rp}{\log \epsilon}. \label{eq:NZ} \eeq Note
that $\kappa_{X}\geq0$.

We also introduce the following relative scaling parameters for the
average signal-to noise-ratio, $\SNR_i$, and interference-to-noise
ratio, $\INR_i$, at the receivers, along the lines of \cite{ETW08}.
\begin{align}
\beta_{i} &\df \frac{\log \SNR_i}{\log \SNR}, \label{eq:hvar1} \\
\alpha_{i} &\df \frac{\log \INR_i}{\log \SNR}. \label{eq:hvar2}
\end{align}
Here $\SNR = P_1 /N_0$. The parameters $\lbr \beta_{1}, \alpha_{1},
\beta_{2}, \alpha_{2} \rbr$ are the relative strength level of the
direct and cross links in dB scale. We will study the scaling
behavior of the rate and reliability with $\SNR$ keeping these
parameters fixed.

We say that a diversity-multiplexing point
$(d_{1},d_{2},r_{1},r_{2})$ is achievable for the two-user
interference channel with parameters $\lbr \beta_{1}, \alpha_{1},
\beta_{2}, \alpha_{2} \rbr$ if there exists a sequence of schemes,
one for each (integer-valued, say) SNR, with rate pair $(R_{1},R_{2})$ given by \beq R_{1}
= r_{1}\log\SNR \quad \textrm{and} \quad R_{2} = r_{2}\log\SNR, \eeq
and probability of error for the two receivers $\prob(\C{E}_1)$ and
$\prob(\C{E}_2)$ respectively, satisfying \beq \prob(\C{E}_1) \leqo
\SNR^{-d_{1}} \quad \textrm{and} \quad \prob(\C{E}_1) \leqo
\SNR^{-d_{2}}. \eeq The set of all achievable diversity-multiplexing
points constitute the fundamental diversity-multiplexing tradeoff
region. By $\mathcal{R}(d_{1},d_{2})$, we will denote the
fundamental multiplexing gain region for given diversity gains
$d_{1}$ and $d_{2}$. To characterize the DMT region, our approach
will be to characterize $\mathcal{R}(d_{1},d_{2})$ for every
diversity gain pair $(d_{1},d_{2})$.

\section{Fundamental DMT Region} \label{sec:main}

The outage formulation provides a natural approach to build
fundamental limits on the DMT. In this
section, we will characterize the fundamental DMT for the two-user
slow fading interference channel by relating it to the compound  interference
channel defined over an appropriate no-outage set.

For given diversity gain requirements $d_{1}$ and $d_{2}$ for the
two users, we characterize the fundamental multiplexing gain region
$\mathcal{R}(d_{1},d_{2})$. We do this by defining outage events
$O_{1}$ and $O_{2}$ for the two users with probability of the order
$\SNR^{-d_{1}}$ and $\SNR^{-d_{2}}$ respectively. These outage
events are defined so that they are largest in the scale of
interest. Then, we use the compound channel result of \cite{RPV08}
to characterize the generalized degree of freedom region of the
compound interference channel defined by the no-outage set, denoted
by $\mathcal{C}(d_{1},d_{2})$.

The main result of this paper  is that the generalized degree of
freedom of the compound interference channel is the fundamental
multiplexing region of the interference channel. Further, the
achievability theorem for the compound interference channel in
\cite{RPV08} ensures that the generalized multi-level
superposition coding scheme with opportunistic decoding of
interference achieves the  fundamental DMT of the interference
channel.

\subsection{Outage Formulation}

We introduce the following change of variables as is done in
\cite{ZT03}.
\begin{align}
{\hat{X}} \df -\frac{\log \lp |X|^{2} \rp}{\log \SNR},
\label{eq:ordRV}
\end{align}
for $X\in\lbr h_{1},g_{1},h_{2},g_{2}\rbr$. $\hat{X}$ is in a way
the measure of the order of the magnitude of the channel gain with
respect to $\SNR$.

\begin{figure}[htb]
\begin{center}
\scalebox{0.3}{\input{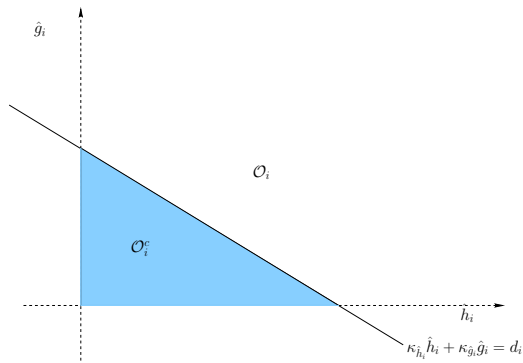}}
\end{center}
\caption{Partition of $(\hat{h}_i,\hat{g}_i)$-space into outage and
no-outage sets} \label{fig-Osets}
\end{figure}

We now define the complement of the outage events for the two
receivers, i.e.~$\hat{\mathcal{O}}^c_{1}$ and
$\hat{\mathcal{O}}^c_{2}$ respectively, in terms of the order random
variables $\hat{X}$'s. The outage and the no-outage sets are
pictorially depicted in Figure \ref{fig-Osets}.
\begin{align}
\hat{\mathcal{O}}_{i}^{c} \df \lbr \lp \hat{h}_i, \hat{g}_i \rp :
\kappa_{h_{i}}\hat{h}_{i}+\kappa_{g_{i}}\hat{g}_{i} <
d_{i},\;\hat{h}_{i},\hat{g}_{i}\geq0  \rbr. \label{eq:outageCset}
\end{align}
The outage events so defined satisfy two important properties as
stated in the following Lemma.

\begin{lemma}\label{lem:outage}
\begin{enumerate}
\item[(i)]
If $E_i^\epsilon \subset \hat{\mathcal{O}}_{i}^{c}$ is an
infinitesimal small square element given by, \beq E_i^\epsilon =
[h^\prime,h^\prime+\epsilon)\times[g^\prime,g^\prime+\epsilon),
\label{eq:Eieps} \eeq then \beq \mathbb{P}(E_i^\epsilon) \geqo
\SNR^{-d_i+\delta}, \eeq for some $\delta>0$.
\item[(ii)]
$\mathbb{P}(\mathcal{\hat{O}}_{i}) \eqo \SNR^{-d_{i}},\qquad i=1,2.$
\end{enumerate}
\end{lemma}

\noindent{\em Proof:} The reader is referred to Appendix
\ref{app:outage} for the details of the proof. The result follows
from the properties \eqref{eq:expTail} and \eqref{eq:NZ} assumed for
the distribution functions. In particular part (ii) follows from
Laplace's method as in \cite{ZT03} and \cite{TV06}. \hfill$\Box$.

We now consider the compound interference channel defined by the set
of channel states $\hat{\A}$, where \beq\hat{\A} =
\hat{\mathcal{O}}_{1}^{c} \times \hat{\mathcal{O}}_{2}^{c}.\eeq We
denote this by $\mathcal{I}(\hat{\A})$. Note that since $\hat{\A}$
is a function of SNR, $\mathcal{I}$ is a family of compound
interference channels indexed by  $\SNR$, with fixed parameters
$\lbr \beta_{1}, \alpha_{1}, \beta_{2}, \alpha_{2} \rbr$ and
$(d_1,d_2)$.

Let $\C{C}(\mathcal{\hat{\A}})$ denote the capacity region of the
compound interference channel $\mathcal{I}(\hat{\A})$. We define the
generalized degree of freedom (g.d.o.f.) for this family of compound
interference channel $\mathcal{I}$ as \beq \mathcal{C}(d_1,d_2) \df
\limsup_{\SNR\rightarrow\infty}
\frac{\C{C}(\mathcal{\hat{\A}})}{\log\SNR}. \eeq Note that this is
an extension of the definition of the g.d.o.f.~of a family of
interference channels in \cite{ETW08} to the case of a family of
compound interference channels.

\subsection{Main Result}

The main result of this paper is the following theorem which relates
the fundamental multiplexing region of the two-user interference
channel $\mathcal{R}(d_{1},d_{2})$ defined in section
\ref{sec:model} to the g.d.o.f.~defined of the family of compound of
interference channel $\mathcal{C}(d_1,d_2)$ defined in the previous
subsection.

\begin{thm}\label{thm:main}
\beq \mathcal{R}(d_{1},d_{2}) = \mathcal{C}(d_{1},d_{2}). \eeq
\end{thm}

\noindent{\em Proof:}

We will first show that $\mathcal{C}(d_{1},d_{2}) \subseteq
\mathcal{R}(d_{1},d_{2})$. Consider $(r_{1},r_{2})\in
\mathcal{C}(d_{1},d_{2})$. This implies that there is a family of
schemes, one for each $\SNR$, that achieves the rate pair $\lp r_{1}
\log \SNR , r_{2} \log \SNR \rp$ for the compound interference
channel $\mathcal{I}(\hat{\A})$. For these schemes the probability
of error at receiver $i=1,2$, can be bounded as
\begin{align}
\prob(\C{E}_{i}) &= \prob(\C{O}_{i})\prob(\C{E}_{i}|\C{O}_{i}) +
\prob(\C{O}^{c}_{i})\prob(\C{E}_{i}|\C{O}^{c}_{i}).
\end{align}
Since $(r_{1},r_{2})\in \mathcal{C}(d_{1},d_{2})$, we have
$\prob(\C{E}_{i}|\C{O}^{c}_{i})\rightarrow 0$. Thus, we can bound
the probability of error as
\begin{align}
\prob(\C{E}_{i}) \leq \prob(\C{O}_{i}) \eqo \SNR ^{-d_{i}}.
\end{align}
Therefore, $(r_{1},r_{2}) \in \mathcal{R}(d_{1},d_{2})$.

Next we will show that $\mathcal{R}(d_{1},d_{2}) \subseteq
\mathcal{C}(d_{1},d_{2})$ Consider $(r_{1},r_{2}) \in
\mathcal{R}(d_{1},d_{2})$. There is a family of schemes, one for
each $\SNR$, with rate pairs $\lp r_{1}\log \SNR,r_{2} \log \SNR
\rp$ and with probability of error at the two receivers,
$\prob(\C{E}_{i}),\; i=1,2$ satisfying \beq \prob(\C{E}_{i}) \leq
\SNR ^{-d_{i}}. \eeq Let $E_i^\epsilon$ denote any infinitesimally
small subset of $\hat{\C{O}}^{c}_{i}$ as defined in
\eqref{eq:Eieps}. From Lemma \ref{lem:outage}, we know that \beq
\prob(E_i^\epsilon) \geqo \SNR ^{-d_{i}+\delta}. \eeq We can lower
bound the probability of error as follows \beq \prob(\C{E}_{i}) \geq
\prob(\C{E}_{i} \cap E_i^\epsilon)  =
\prob(E_i^\epsilon)\prob(\C{E}_{i}|E_i^\epsilon). \eeq Therefore
\beq \prob(\C{E}_{i}|E_i^\epsilon) \leq
\frac{\prob(\C{E}_{i})}{\prob(E_i^\epsilon)} \leq \SNR^{-\delta}.
\eeq As $\SNR\rightarrow\infty$, $P(\C{E}_{i}|E_i)\rightarrow 0$.
Therefore, $(r_{1}, r_{2}) \in \C{C}(d_1,d_2)$. And hence
$\mathcal{R}(d_{1},d_{2}) \subseteq \C{C}(d_{1},d_{2})$.
\hfill$\Box$

\subsection{Achievability of the Fundamental DMT Region}

In \cite{RPV08}, a coding theorem for the compound two-user Gaussian
interference channel was given. The scheme used was a multi-level
superposition coding scheme with opportunistic decoding of
interference. This scheme is a generalization of the two-level
superposition coding scheme, which is also called the Han-Kobayashi
scheme. From theorem \ref{thm:main} and the coding theorem in
\cite{RPV08} the following result follows.

\begin{thm}
 The multi-level superposition coding scheme with opportunistic decoding of interference is approximately universal for the slow-fading interference channel, i.e.~it achieves the fundamental DMT region of the scalar two-user interference channel.
\end{thm}

\noindent{\em Proof:} It was shown in \cite{RPV08} that multi-level superposition coding scheme
achieves within 1 bit/symbol of the capacity region of the {\em
finite} state compound interference channel. The finite state
assumption required that the set of states defining the compound
channel have finite cardinality. If the number of states is $N$,
then the scheme essentially has $N+1$ independent data streams
superposed together. Depending on the channel state, a suitable
number of streams from the interfering user is decoded. In our
problem, the set $\A$ is a continuum of states and hence the
compound interference channel defined by the set $\A$ has infinite
states.

The compound interference channel result of \cite{RPV08} for the
finite state can be easily extended to the infinite state by doing a
finite quantization of the infinite state set. We can then use a
finite-level superposition coding scheme corresponding to the
quantized finite state compound channel. We can get as close as
desired to within 1 bit/symbol of the capacity region by taking a
fine enough quantization.

In the limiting case, as SNR goes to infinity, the constant 1
bit/symbol gap disappears and the superposition coding scheme
achieves the g.d.o.f.~$\C{C}(d_{1},d_{2})$ and due to Theorem
\ref{thm:main} achieves $\C{R}(d_{1},d_{2})$. \hfill$\Box$.

\section{Han-Kobayashi Scheme and Symmetric DMT} \label{sec:symCase}

In the previous section we showed that the multi-level superposition
coding scheme with opportunistic decoding of interference achieves
the DMT of the two-user interference channel, universally for all
fading channel statistics. However, an explicit characterization of
the DMT region is elusive. This is due to the complexity in
explicitly finding the g.d.o.f.~region for the compound interference
channel achieved by the multi-level superposition coding scheme as
seen in \cite{RPV08}. 

The two-level version of this scheme is the
familiar Han-Kobayashi scheme. It is an interesting question to ask
as to how far from optimal is this scheme with respect to the DMT.
In Section \ref{subsec:symCaseHK}, we answer this question for the
symmetric case, where we impose identical rate and diversity
requirements for both the users. It turns out that the Han-Kobayashi
scheme is DMT optimal in many regimes. We show this by comparing its
performance to an outer bound derived from considering the worst
case outer bound of the corresponding compound interference channel
in Section \ref{subsec:symCaseWCOB}. However, in Section
\ref{subsec:symCaseBHK} we show with an example that both the
Han-Kobayashi scheme and the worst case outer bound are not tight
for all regimes.

\subsection{DMT Achieved by the Han-Kobayashi Scheme} 
\label{subsec:symCaseHK}
The focus of this section is on the DMT performance of the
Han-Kobayashi scheme. We analyze its performance explicitly, but
in the specific context below:
\begin{itemize}
\item We suppose the statistics of the channel gains to be symmetric
with respect to the two users i.e.,
\begin{align}
\beta_{1}=\beta_2 = 1,\qquad \alpha_{1}=\alpha_{2} = \alpha.
\end{align}
\item We suppose that the near-zero behavior of the
distribution function to be linear (example: Rayleigh distribution),
i.e.~$\kappa_X =1$, for $X=\lbr h_1,g_1,h_2,g_2 \rbr$. This implies that
\beq  \mathbb{P}\lp\frac{|X|^{2}}{E[|X|^{2}]}<\epsilon\rp \approx
\epsilon, \eeq for small $\epsilon>0$.
\item We only consider
symmetric diversity and rate requirements for the two receivers,
i.e.,
\begin{align}
d_{1} = d_{2} =d, \qquad r_{1} = r_{2} = r.
\end{align}
\end{itemize}

The Han-Kobayashi scheme  is a
two-level superposition scheme and achieves the generalized degree of freedom
region of the interference channel \cite{ETW08}.
 The  scheme can be succinctly described as
the following. Each user splits its message into two parts - {\em public}
and {\em private} - and superposition-codes them. The public is decoded
by both the receivers, while the private is decoded only by the
intended receiver. Gaussian code books with rate $s_{i} \log \SNR$
and $t_{i} \log \SNR$ are used for both the private and public
messages respectively. The actual codeword transmitted is the sum of
the public and private codewords. Therefore, the rate achieved for
each user is $r_{i} \log \SNR$, where \beq r_{i} = s_{i} + t_{i}.
\eeq

An important parameter of this scheme is the power split between the
public and private streams. Let $\rho_{v_{i}}$ denote the fraction
of power allocated to the private of user $i$. Since we are
concerned with high SNR approximations, we will let \beq
\rho_{v_{i}} \eqo \SNR^{-v_{i}}, \eeq where $v_{i}>0$. The fraction
of power allocated to the public will be given by \beqa
\rho_{u_{i}} &=& 1-\rho_{v_{i}} \nn\\
&\eqo& 1 - \SNR^{-v_{i}} \nn\\
&\eqo& 1. \eeqa The parameter $v_{i}$, therefore, determines the
power split for user $i$.

Consider the case when there is no fading; so, the channel gains
are the mean values, i.e.~unity. The interference level (in the
relative dB scale) is $\alpha$. In \cite{ETW08} it was shown that a
good choice for the power split is such that the private stream
appears at noise level at the interfering receiver i.e.,
\beqa &\rho_{v_{i}} \SNR^{\alpha} &= 1 \nn\\
 & v_{i} &= \alpha.\eeqa Suppose the channel cross
gains $|g_{i}|^{2}$ take values $\SNR^{-\hat{g_{i}}}$, for $i=1,2$.
Then the effective interference level for the two links are
$\alpha-\hat{g}_{i}, i=1,2$. Correspondingly, the power split for the two
users is given by $\alpha - \hat{g}_{2}$ and $\alpha-\hat{g}_1$,
respectively.

It is interesting to note that the power split does not depend on the level
of the direct links. The direct links only limit the rates. For our
compound channel problem, as the cross gain takes values in the
no-outage set, the interference level takes values in the range
$(\alpha-d,\alpha)$ for each link independently. We need to find a
fixed scheme with optimal power-split and suitable rates for the
public and private messages, that works for the compound channel. This is
a routine optimization problem and the results of this
calculation (done in Appendix \ref{app:HKscheme}) are summarized in
Table \ref{tab:HK}. Figure~\ref{fig-symDMT} illustrates the optimal tradeoff
curve that can be achieved by the Han-Kobayashi scheme for different
values of $\alpha$.

\begin{table}[ht]
\caption{DMT achieved by the Han-Kobayashi scheme}\label{tab:HK}
\begin{tabular}{|c|c|c|c|}
\hline
$\alpha$ & $d$ & $v$ & $r_{\mathrm{HK}}(\alpha,d)$ \\
\hline\hline
$\alpha\geq1$& $1\geq d\geq 0 $ &  (All public) & $(1-d) \wedge \frac{1-d+\alpha}{4} $ \\
\hline
$1>\alpha\geq0$ & $1 \geq d > 1-  \alpha$ & (All public) &  $  (1-d) \wedge \frac{1-d+\alpha}{4} $ \\
\hline
$1>\alpha\geq\frac{2}{3}$ & $1-\alpha \geq d \geq 0$ & $\alpha$ & $ 1-\frac{\alpha}{2}-d$\\
\hline
$\frac{2}{3} > \alpha \geq \frac{5}{8}$ & $1-\alpha \geq d > 5\alpha-3$ & $\frac{3\alpha+d-1}{2}$ &$\frac{3-\alpha-3d}{4}$\\
& $5\alpha-3 \geq d \geq 0$ & $\alpha$ &$(\alpha-d)$\\
\hline
$\frac{5}{8} > \alpha \geq 0$ & $1-\alpha \geq d > 1-\frac{7\alpha}{5}$ & $\frac{3\alpha+d-1}{2}$ &$\frac{3-\alpha-3d}{4}$\\
& $1-\frac{7\alpha}{5} \geq d > (\alpha-\frac{1}{2}) \vee (1-2\alpha)$ & $\frac{1+\alpha-d}{3}$ &$\frac{1+\alpha-d}{3}$\\
& $(\alpha-\frac{1}{2}) \vee (1-2\alpha) \geq d \geq 0$ & $\alpha$ &$(\alpha-d)\vee (1-\alpha-d)$\\
\hline
\end{tabular}
\end{table}

\subsection{Comparison to the Worst Case Outer Bound} \label{subsec:symCaseWCOB}

\begin{figure}[p]
\subfloat[$1 \leq
\alpha$]{\label{fig-dmta125}\scalebox{0.25}{\includegraphics{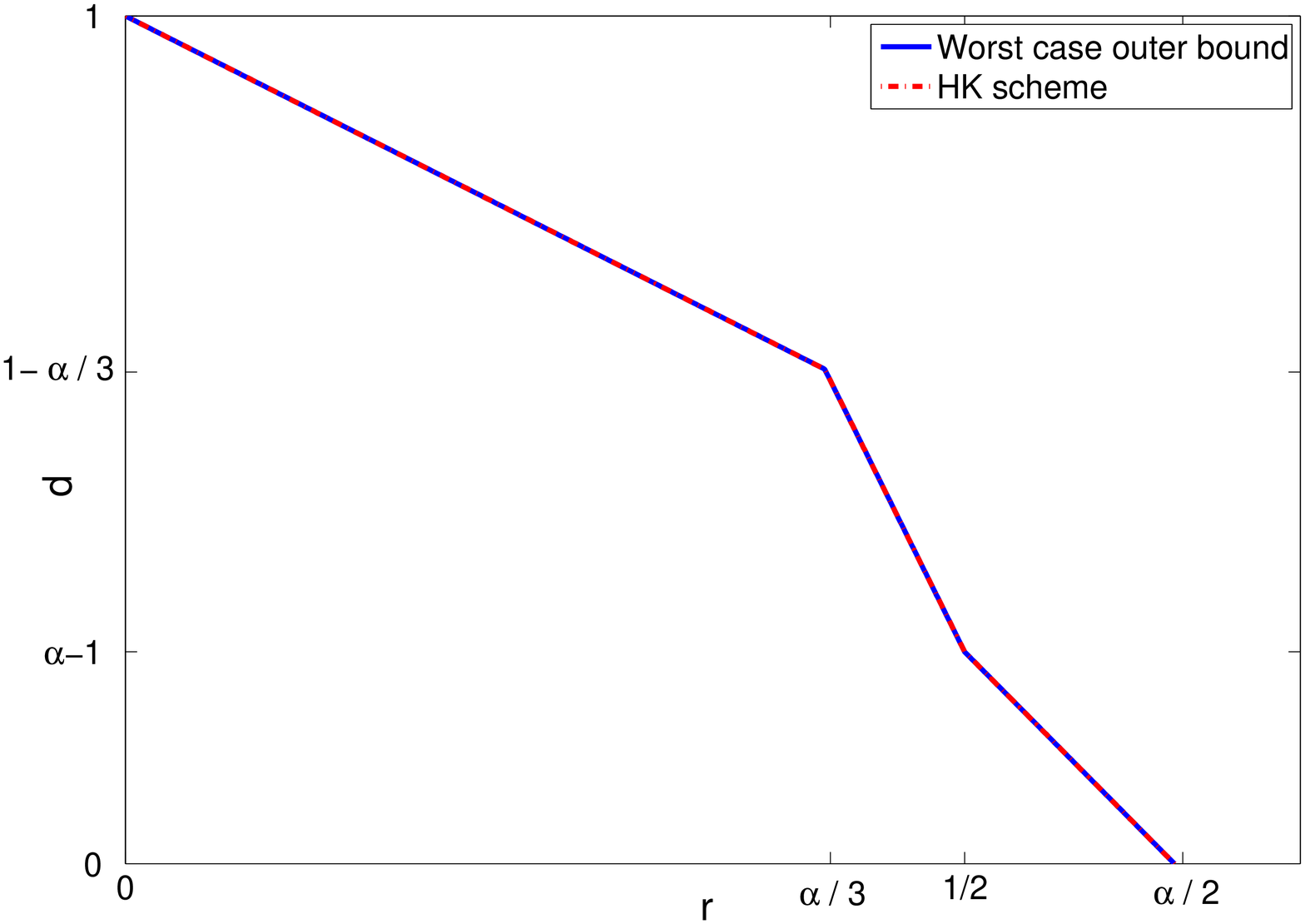}}}
\subfloat[$2/3 \leq \alpha \leq 1$]{\label{fig-dmta067}\scalebox{0.25}{\includegraphics{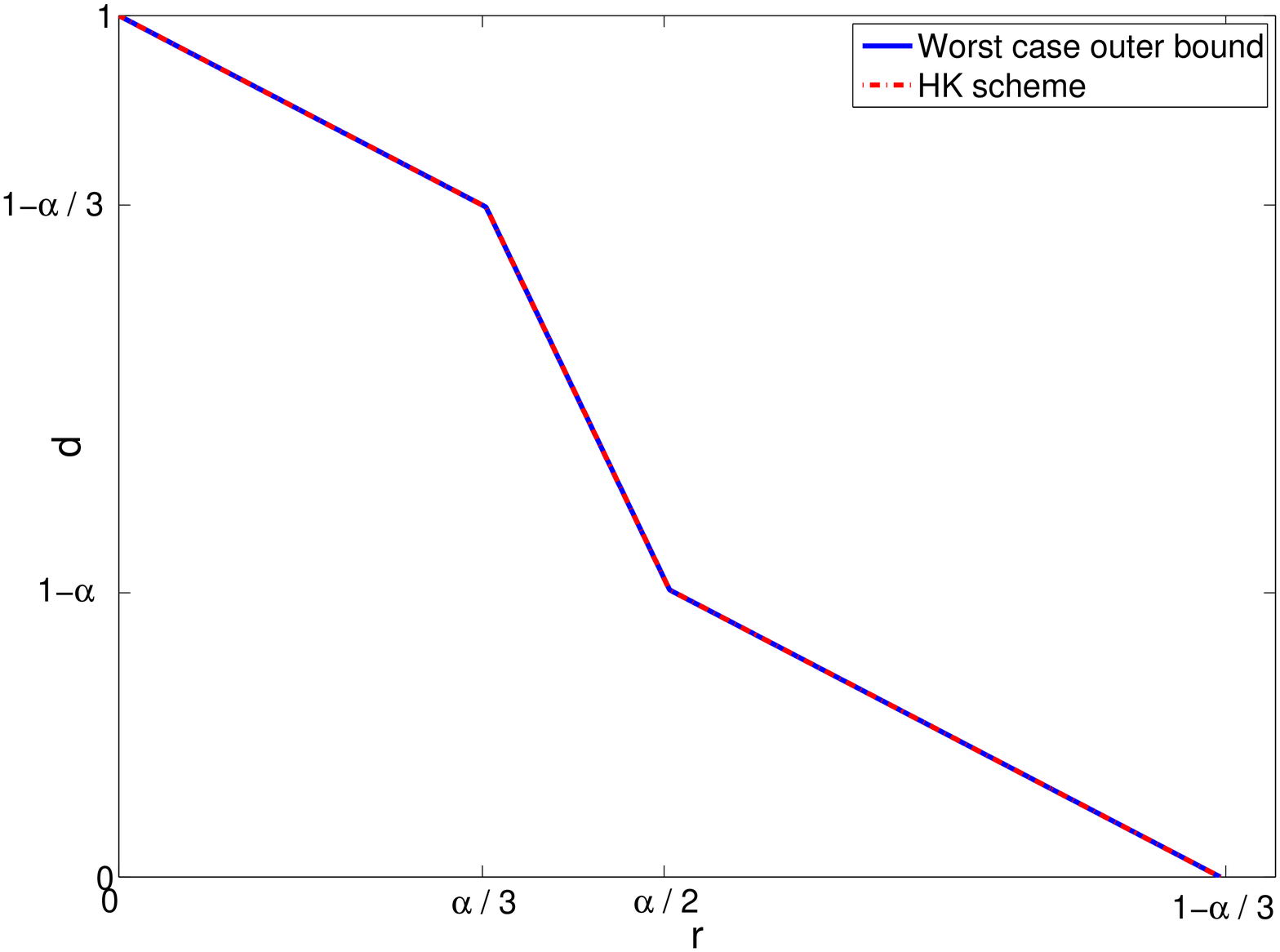}}}\\
\subfloat[$5/8 \leq \alpha \leq
2/3$]{\label{fig-dmta065}\scalebox{0.25}{\includegraphics{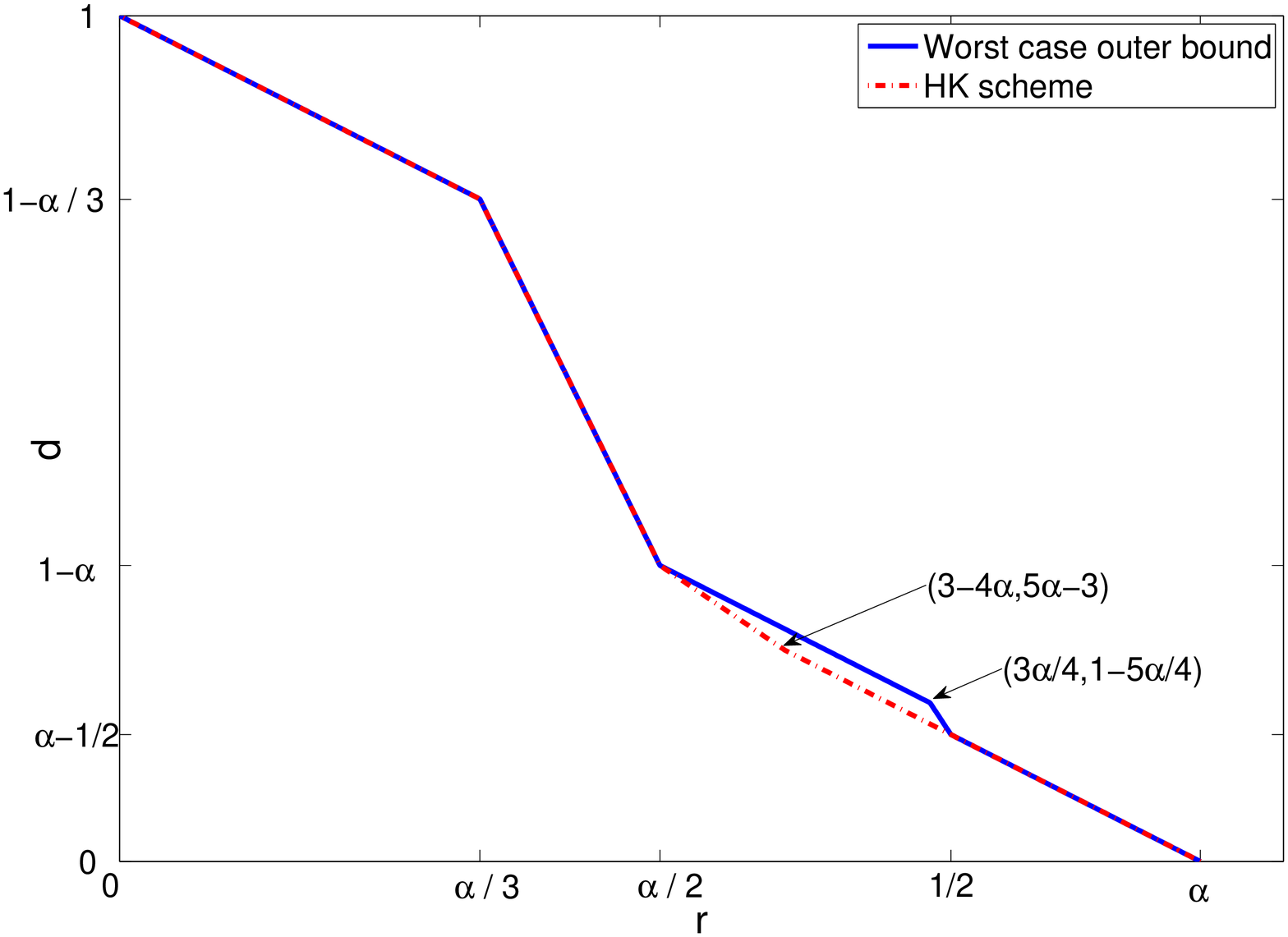}}}
\subfloat[$1/2 \leq \alpha \leq 5/8$]{\label{fig-dmta06}\scalebox{0.25}{\includegraphics{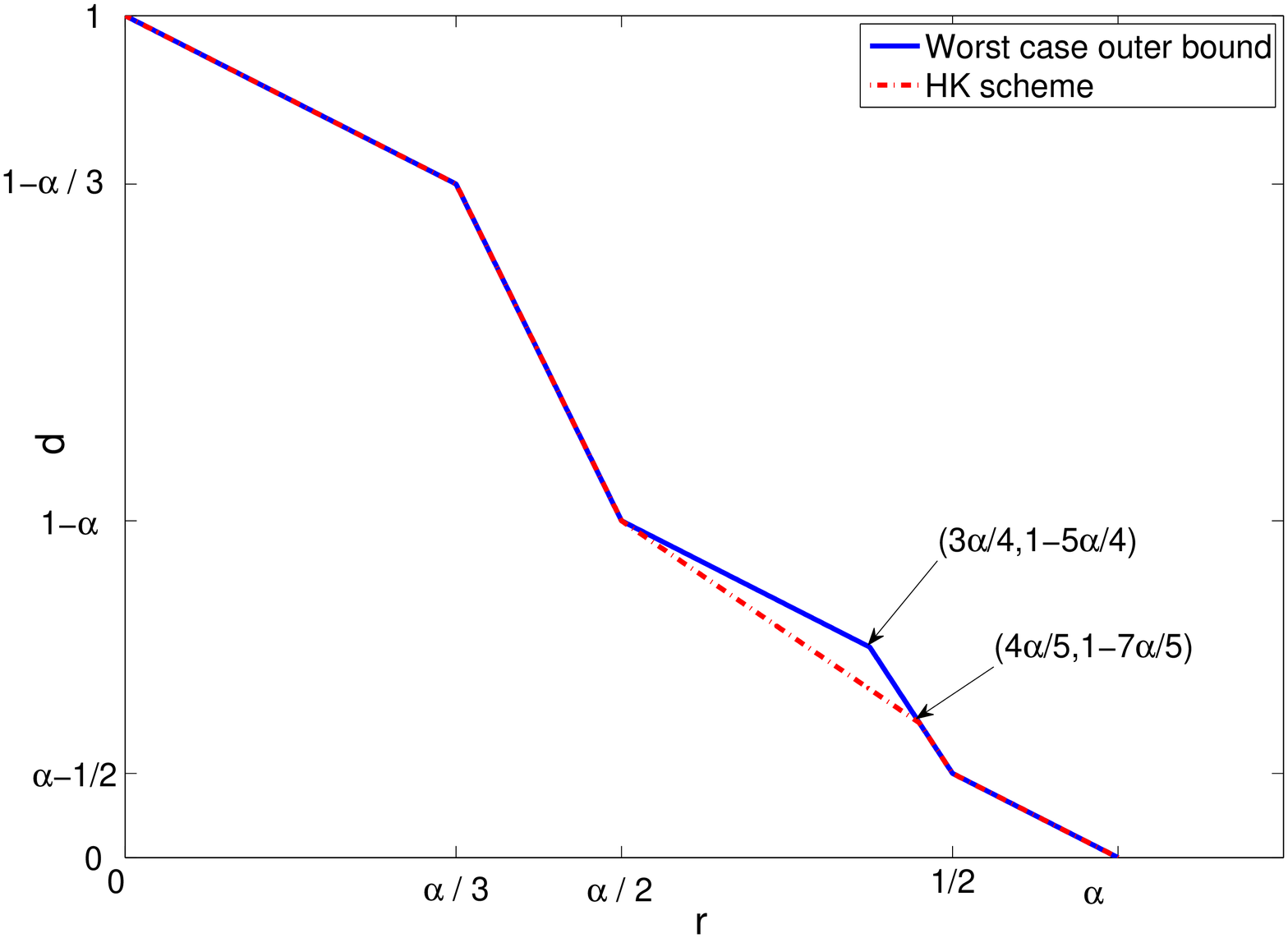}}}\\
\begin{center}\subfloat[$0 \leq \alpha \leq 1/2$]{\label{fig-dmta025}\scalebox{0.30}{\includegraphics{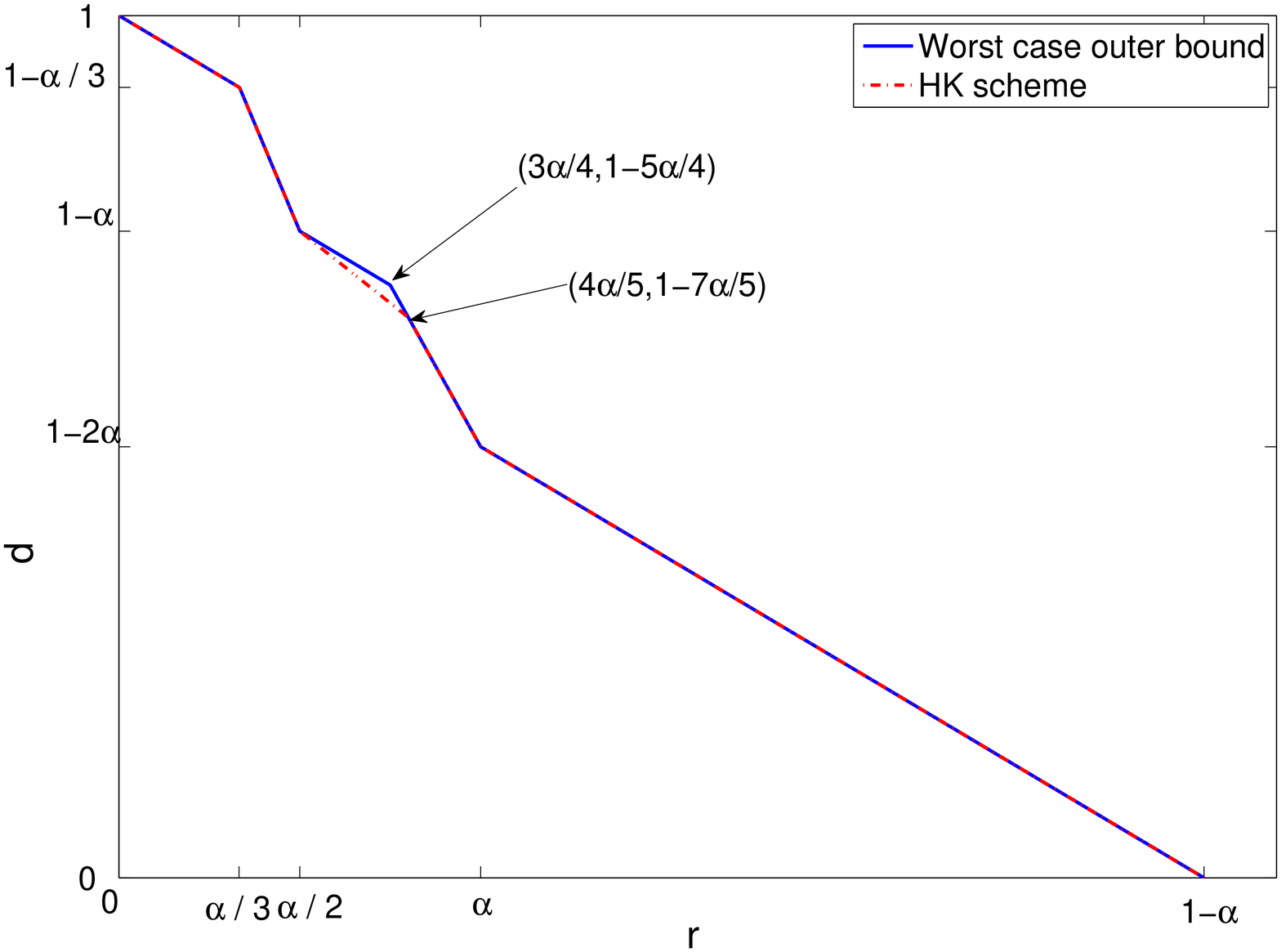}}}\end{center}
\caption{Symmetric DMT} \label{fig-symDMT}
\end{figure}

To evaluate the performance of the Han-Kobayashi scheme we  use
a  simple outer bound to the DMT, obtained by using the worst case
outer bound of the compound channel. This is equivalent to supposing
that the transmitter also has channel information. We know that the
capacity of the compound channel cannot be larger than the capacity
of the worst channel in the set, i.e., \beq C(\hat{\A}) \leq
\text{min}_{\hat{a}\in\hat{\A}}\; C(\hat{a}). \eeq Here
$C(\hat{\A})$ is the symmetric d.o.f.~of the compound channel and
$C(\hat{a})$ is the symmetric d.o.f.~of the two-user interference
channel with channel coefficients $\hat{a}=(\hat{h}_{1},\hat{g}_{1},
\hat{h}_{2}, \hat{g}_{2})$ and can be easily computed \cite{ETW08}.
The detailed analysis can be found in Appendix \ref{app:WCob}. The
worst case outer bound is also plotted for different values of
$\alpha$ in Fig.~\ref{fig-symDMT}

Note that when the diversity gain needed is larger, we need a coding scheme that is resilient to a larger range of fading.
We can summarize our understanding
by breaking the diversity gain into three regimes. \begin{itemize}
\item {\em Large diversity regime:} When the diversity gain is large enough $(d>(1-\alpha)^{+})$, the dominant error event  is the direct link fading so much, so as to drive the interference channel into the strong and very strong interference regime. Therefore decoding the complete message from both the users at both the receivers is DMT optimal in this regime.  
\item {\em Small diversity regime:} We define this to be the regime where the Han-Kobayashi scheme with the power split corresponding to the mean level of interference $(v=\alpha)$ is DMT optimal. This is true when the diversity gain is small enough (which depends on $\alpha$ as seen from Fig.~\ref{fig-symDMT} and Table \ref{tab:HK}). Since the fading range over which our coding scheme must work is small, the perturbation in the strength of the channel links from the mean level is small and a scheme designed for the mean level still remains optimal in the DMT sense. 
\item {\em Intermediate diversity regime:} This is the regime which is complement to the first two regimes. Our optimization suggests what power-split level to choose for the Han-Kobayashi scheme. While this choice is DMT optimal for some part of the regime, in general it is not. This is due to the non-monotonic behavior of the symmetric rate versus the cross link strength as seen from the well known `W-shaped' curve in \cite{ETW08}.
\end{itemize}

\subsection{Improving the HK scheme and the Outer Bound} \label{subsec:symCaseBHK}

The previous section shows that the Han-Kobayashi scheme does not
meet the worst case outer bound in many regimes. In the rest of this section,
we see that {\em both} the inner and outer bounds can be improved.
We will do this by considering the specific case when
\beqa
\alpha &=& 1/2 \\
d &=& 1-5\alpha/4 = 3/8.
\eeqa
From Figure \ref{fig-dmta06}, we see that \beq
r_\textrm{HK}(1/2,3/8) = 11/32 = 0.34375,\eeq whereas \beq
r_\textrm{out}(1/2,3/8) = 3/8 =  0.375. \eeq

\subsubsection{Beyond Han-Kobayashi Scheme}

In \cite{RPV08}, we generalized the Han-Kobayashi scheme to a
multi-level superposition coding scheme. In particular, we will now
consider a three-level superposition coding scheme. Accordingly,
each user splits its message into three parts with power splits
given by, \beq \rho_{v_{i1}} \eqo \SNR^{-v_{1}},\quad \rho_{v_{i2}}
\eqo \SNR^{-v_{2}},\;\textrm{and}\quad \rho_{v_{i3}} \eqo
1,\qquad\textrm{for }i=1,2. \eeq The receiver decodes one or two
messages from the interfering user depending on the interference
level. Accordingly, we will define a threshold $\gamma$ such that,
when $\hat{g}_i \geq \gamma$ the receiver decodes two messages, and
when $\hat{g}_i < \gamma$ the receiver decodes one message. The
resulting expressions describing the rates achievable by this scheme
are prohibitively complex to describe. And hence, we will just
describe our results. Optimizing the values for $v_1, v_2$ and
$\gamma$ to achieve the maximum symmetric rate gives, \beq
r(1/2,3/8) = 55/152 = 0.361842105, \eeq when, \beq v_1 = 73/152,
\quad v_2 = 49/152,\;\textrm{and}\quad \gamma = 56/152. \eeq

\subsubsection{Beyond the worst-case outer bound}

We can obtain a better outer bound to the DMT by obtaining a better
outer bound to the capacity of the compound channel described by the
$\hat{\A}$. We can do this by taking a finite-state quantization of
the set $\hat{\A}$, say $\hat{\A}_Q$. In \cite{RPV08}, we described
the capacity of the finite state compound interference channel. Note
that this capacity region was obtained as a two-dimensional
projection of many higher dimensional polytopes and was complicated
to describe. However, numerically, given fixed values of the states,
the symmetric capacity can be obtained easily by a linear program.

In our analysis, we used a uniform quantization given by \beq
\hat{g}_{ik} = \alpha - \frac{kd}{Q-1},\;\textrm{and}\quad
\hat{h}_{ik} = d - \hat{g}_{ik},\qquad 0\leq k\leq Q-1,\; i=1,2.
\eeq For $Q=16$, we can get \beq r_\textrm{out}(1/2,3/8) = 0.3667.
\eeq

\section{Important Instances}\label{sec:spCases}
In our discussion so far we have considered all four links in the interference
channel to be potentially faded. In this section, we consider two
importance instances of the interference channel which do not share this
characteristic. In Section \ref{subsec:ZIFC} we will consider the case of
Z-interference channel, where only one of the user is interfering
with the other. In Section \ref{subsec:CLfade} we will consider the
case where only the cross links are fading and the direct links are
fixed. Both these cases share an interesting result: we can have
a closed form characterization of the optimal DMT and, furthermore, the
Han-Kobayashi scheme achieves it.

\subsection{Z-Interference Channel}\label{subsec:ZIFC}

\begin{figure}[h]
\begin{center}
\scalebox{0.65}{\input{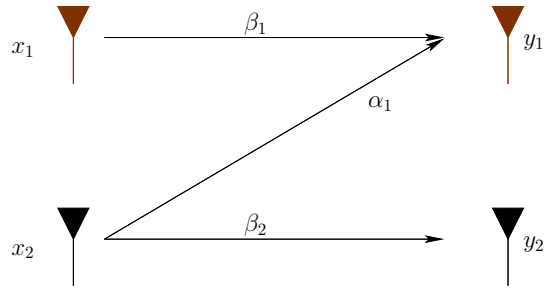}}
\end{center}
\caption{The slow fading two-user Gaussian Z-interference channel. }
\label{fig-ZICmodel}
\end{figure}

The Z-Interference channel is illustrated in Figure \ref{fig-ZICmodel}.
The figure shows the average relative strength of the links
$\beta_{1}, \beta_{2}$ and $\alpha_{1}$ as given by \eqref{eq:hvar1}
and \eqref{eq:hvar2}. Note that for the Z-Interference channel the
channel gain coefficient $g_{2}$ is zero; this implies that
$\alpha_{2}=-\infty$. For simplicity of analysis, we will assume the
near-zero behavior of the distribution function to be linear, i.e.
$\kappa_{X} =1$ for $X=\lbr h_{1},g_{1},h_{2} \rbr$ (example: Rayleigh
distribution).

\begin{thm} \label{thm:DMTZIFC}
The fundamental multiplexing gain region $\C{R}(d,d)$ for
the Raleigh fading two-user Gaussian Z-interference channel with $\beta_{1}=\beta_{2}=1$ and $\alpha_{1}=\alpha$ is given by
\begin{align}
\C{R}(d,d) = \lbr r_{1}\right. &\leq (1-d)^{+}, \nn\\
r_{2} &\leq (1-d)^{+}, \nn\\
r_{1}+r_{2} &\leq\left.
\lp1-d\vee\alpha-d\vee
\frac{1-d+\alpha}{2}\rp+(1-\alpha-d)^{+}\rbr.
\end{align}
Further, a simple Han-Kobayashi scheme achieves this fundamental DMT
region.
\end{thm}
\noindent{\em Proof:} The details of the proof are in Appendix
\ref{app:DMTZIFC}. The idea is to characterize the DMT region
achievable by a simple Han-Kobayashi scheme with power split
$v_{1}=0$ and $v_{2}=\alpha_{1}$. It turns out that this meets the
simple worst case outer bound. \hfill$\Box$

\subsection{Fading Interfering Links}\label{subsec:CLfade}

In many practical settings, a wireless link (between a single source
and destination) is power controlled so as to effectively have no
effect of fading.  We turn to such a
scenario now: the direct link gains $h_1$ and $h_2$ are fixed, while the
interfering links are susceptible to channel fading. This is equivalent to
assuming that \beq \kappa_{h_i}=0, \quad i=1,2.\eeq  For simplicity, we will also assume the
average relative strength of the links is symmetric and is given by
\begin{align}
\beta_{1}=\beta_2 = 1,\qquad \alpha_{1}=\alpha_{2} = \alpha.
\end{align}
As usual we suppose that the near-zero behavior of the
distribution function of the cross gains $X=\lbr g_{1},g_{2}\rbr $
to be linear, i.e., $\kappa_{X} =1$ (example: Rayleigh distribution).

\begin{thm}
The symmetric DMT of the interference channel with only cross links
fading is given by \beq r(d) = 1 \wedge \lp \frac{1}{2} \vee
1-\alpha \vee \alpha-d \rp \wedge \lp \frac{1}{2} \vee
1-\frac{\alpha}{2} \vee \frac{\alpha-d}{2} \rp. \label{eq:DMTspcase}
\eeq
\end{thm}
\noindent{\em Proof:} The no-outage set is now given by \beq
\hat{\A} = \lbr 0 \leq \hat{g}_1 \leq d_1 \rbr \times \lbr 0 \leq
\hat{g}_2 \leq d_2 \rbr. \eeq We obtain the worst case outer bound
by following the analysis similar to that in Appendix \ref{app:WCob}
and using \eqref{eq:rout} (Note that for our current setup
$\hat{h}_{1}=\hat{h}_{2}=0$). It can be shown that the worst case
outer bound is \eqref{eq:DMTspcase}.

Turning to achievable schemes: we first note that the simple scheme of
orthogonalizing the two users, so that they don't interfere with each
other, achieves a multiplexing gain of $1/2$ to support arbitrary
diversity. This is possible because in this model the direct links
are not fading. This scheme meets the outer-bound for $d \geq \alpha
-1$, when $\alpha > 1$, and for $d \geq \alpha -1/2$ when $\alpha
\leq 1$.

Another simple scheme is to treat interference as noise and it can
be shown that for $\alpha \leq 1/2$, it meets the optimal
multiplexing gain of $1-\alpha$ for arbitrary diversity gains.

For the rest of the cases, we can characterize the symmetric DMT
achieved by the Han-Kobayashi scheme with power splits given by $v_1
= v_2 =\alpha$. Using \eqref{eq:rsym}, we can show that this scheme
achieves the symmetric DMT given by \beq r(d) = 1 \wedge \lp
1-\alpha \vee \alpha-d \rp \wedge \lp \frac{1}{2} \vee
1-\frac{\alpha}{2} \vee \frac{\alpha-d}{2} \rp. \eeq Hence this
scheme meets the outer bound for the rest of the cases. \hfill$\Box$

If we relax the assumption of the symmetric channel and symmetric
rate-diversity requirements, the analysis is no longer simple. Our
guess is that for this case, the two-split Han-Kobayashi scheme is
no longer optimal. We may need to use generalized schemes with more
than two-splits and also obtain better outer bounds.

\appendix

\section{Proof of Lemma \ref{lem:outage}} \label{app:outage}

\noindent{\em Proof of part (i):} If $E_i^\epsilon \subset
\hat{\mathcal{O}}_{i}^{c}$, where \beq E_i^\epsilon =
[h^\prime,h^\prime+\epsilon)\times[g^\prime,g^\prime+\epsilon),\eeq
then $\kappa_{\hat{h}}h^{\prime}+\kappa_{\hat{g}}g^{\prime} \leq
d_{i}-\delta$ for some $\delta>0$. Further
\begin{align}
 \mathbb{P}(E_i^\epsilon) &= \prob \lp [h^\prime,h^\prime+\epsilon)\times[g^\prime,g^\prime+\epsilon) \rp \nn\\
 &= \prob\lp[h^\prime,h^\prime+\epsilon)\rp\prob\lp[g^\prime,g^\prime+\epsilon)\rp \nn\\
 &\eqo \SNR^{-\kappa_{\hat{h}}h^{\prime}} \SNR^{-\kappa_{\hat{g}}g^{\prime}} \nn\\
 &= \SNR^{-(\kappa_{\hat{h}}h^{\prime}+\kappa_{\hat{g}}g^{\prime})} \nn\\
 &\geq \SNR^{-d_i+\delta},
\end{align}
\hfill$\Box$

\noindent{\em Proof of part (ii):} Note that $\hat{\C{O}}_{i}$
corresponds to the unshaded region in Figure \ref{fig-Osets}. To begin with,
observe that the region where $\hat{h}_{i}$ or $\hat{g}_{i}$
are negative do not contribute to the probability in the scale of
interest. This follows from the exponential tail property of the
distribution function \eqref{eq:expTail}. More precisely for any
arbitrary small $\epsilon>0$ \beq \prob(\hat{h}_{i}\leq -\epsilon) =
\prob(|h_{i}|^{2}) \geq \SNR^{\epsilon}) \eqo 0. \eeq Similarly
$\prob(\hat{g}_{i}\leq -\epsilon) \eqo 0$. Therefore \beq
\prob(\hat{\C{O}}_{i}) \eqo \prob(\hat{\C{O}}_{i}^{\prime}),\eeq
where
\begin{align}
\hat{\C{O}}_{i}^{\prime} &= \hat{\C{O}}_{i} \cap \lbr \real^{+}\times\real^{+} \rbr \nn\\
& = \lbr \kappa_{h_{i}}\hat{h}_{i}+\kappa_{g_{i}}\hat{g}_{i} \geq
d_{i},\;\hat{h}_{i} \geq 0,\;\hat{g}_{i} \geq 0 \rbr.
\end{align}

For $X\in\lbr h_{1},g_{1},h_{2},g_{2}\rbr$ and assuming that the
density function $f_{|X|^{2}}(x)$ exists for small enough $x$, from
\eqref{eq:NZ} it follows that \beq f_{|X|^{2}}(x) \eqo
x^{(\kappa_{X}-1)}, \eeq where the $\eqo$ is in the limit as
$x\rightarrow 0$. If $\hat{X}$ is the change of variables according
to \eqref{eq:ordRV}, then it can easily be shown that \beq
f_{\hat{X}}(x) \eqo \SNR^{-\kappa_{\hat{X}}x}, \quad \forall x \geq
0. \eeq If we let $I=[0,\gamma)^{2}$, where $\gamma>
(d_{i}/\kappa_{h_{i}} \vee d_{i}/\kappa_{g_{i}})$. Then
\begin{align}
\prob (\hat{\C{O}}_{i}^{\prime}) & = \prob (\hat{\C{O}}_{i}^{\prime}\cap I) + \prob (I^{c}) \nn \\
& < \int_{\hat{\C{O}}_{i}^{\prime}\cap I} f_{\hat{h}_{i}}(x) f_{\hat{g}_{i}}(y)\;dx\;dy + \prob \lp [\gamma,\infty] \times [0,\infty] \rp + \prob \lp [0,\infty] \times [\gamma,\infty] \rp \nn\\
& \eqo \int_{\hat{\C{O}}_{i}^{\prime}\cap I} \SNR^{-(\kappa_{\hat{h}_{i}}x+\kappa_{\hat{g}_{i}}y)}\;dx\;dy + \SNR^{-\gamma\kappa_{\hat{h}_{i}}} + \SNR^{-\gamma\kappa_{\hat{g}_{i}}} \nn\\
& < \gamma^{2} \SNR^{-d_{i}} + \SNR^{-\gamma\kappa_{\hat{h}_{i}}} + \SNR^{-\gamma\kappa_{\hat{g}_{i}}} \nn\\
& \eqo \SNR^{-d_{i}}.
\end{align}
Therefore, \beq \prob(\hat{\C{O}}_{i}) \eqo
\prob(\hat{\C{O}}_{i}^{\prime}) \leqo \SNR^{-d_{i}}. \eeq Since
$\lbr [d_{i},\infty] \times [0,\infty] \rbr \subset
\hat{\C{O}}_{i}^{\prime} \subset$ we have, \beq
\prob(\hat{\C{O}}_{i}^{\prime}) \geqo \SNR^{-d_{i}}. \eeq Thus we conclude
\beq \prob(\hat{\C{O}}_{i}) \eqo \SNR^{-d_{i}}. \eeq
\hfill$\Box$.

\section{DMT achieved by the Han-Kobayashi scheme}\label{app:HKscheme}

We characterize the performance of the Han-Kobayashi scheme for the
symmetric DMT by characterizing its performance over the compound
channel given by the no-outage set. The no-outage set for a
symmetric diversity gain of $d$ is given by \beq \hat{\A} = \lbr
\hat{h}_1 + \hat{g}_1 \leq d \rbr \times \lbr \hat{h}_2 + \hat{g}_2
\leq d \rbr, \label{eq:A} \eeq where \beq
(|h_{1}|^{2},|g_{1}|^{2},|h_{2}|^{2},|g_{2}|^{2}) =
(\SNR^{-\hat{h}_{1}},\SNR^{-\hat{g}_{1}},\SNR^{-\hat{h}_{2}},\SNR^{-\hat{g}_{2}}),
\eeq

We use a simple version of the Han-Kobayashi scheme that was shown
to achieve the symmetric capacity within 1 bit per symbol of the
interference channel in \cite{ETW08}. For a given block length $n$
transmitter $i$ chooses a private message from codebook
$\C{C}^v_{i,n}$ and a public message from the codebook
$\C{C}^u_{i,n}$. The codebooks are generated by using i.i.d.\ random
Gaussian variables with variances $\rho_{v_i} = \SNR^{-v_i}$ and
$\rho_{u_i} = 1-\rho_{v_i}$ respectively. The codebooks have rate
$s_i \log\SNR$ and $t_i \log\SNR$ respectively. After selecting the
corresponding codewords user $i$ transmits the signal $\mathbf{x}_i
= P_i (\mathbf{c}^u_i + \mathbf{c}^v_i)$, where $P_i$ is the power
constraint at the transmitter.

Each receiver performs joint decoding of both the  messages from its
transmitter and the public message from the other transmitter. The
decoding of the private message at receiver 1 is successful if,
\beqa
s_1 \log\SNR &\leq& \log \lp 1+ \frac{\rho_{v_1}|h_1|^2\SNR}{1+\rho_{v_2}|g_1|^2\SNR^{\alpha}} \rp \\
&\eqo& \lp 1-\hat{h}_1 - v_1 - (\alpha-v_2-\hat{g}_1)^+ \rp^+ \log
\SNR. \eeqa Therefore, we can write \beq s_1 \leq \lp 1-\hat{h}_1 -
v_1 - (\alpha-v_2-\hat{g}_1)^+ \rp^+ . \eeq Note that since the
channel state can take any value from the no-outage set, we need to
take the minimum over $\hat{\A}$. Therefore \beq s_1 \leq
\mathrm{min}_{\hat{\A}} \lp 1-\hat{h}_1 - v_1 - (\alpha -
v_2-\hat{g}_1)^+ \rp^+ . \eeq

Similarly, we  obtain the other constraints on the sum of
sub-rates by considering the decoding at receivers 1 and 2.
Analogous to the analysis in \cite{CMG08}, we can show that the
following constraints are the only ones that matter: \beqa
s_1 &\leq& a_{11} \df \mathrm{min}_{\hat{\A}} \lp 1-\hat{h}_1 -  v_1 - (\alpha-v_2-\hat{g}_1)^+ \rp^+ \label{eq:hkrates1} \\
t_1+s_1 &\leq& a_{12} \df \mathrm{min}_{\hat{\A}} \lp 1-\hat{h}_1 - (\alpha-v_2-\hat{g}_1)^+ \rp^+ \\
t_2+s_1 &\leq& a_{13} \df \mathrm{min}_{\hat{\A}} \lp ( 1-\hat{h}_1 - v_1 \vee \alpha - \hat{g}_1 ) - (\alpha-v_2-\hat{g}_1)^+ \rp^+ \\
t_2+t_1+s_1 & \leq& a_{14} \df \mathrm{min}_{\hat{\A}} \lp
(1-\hat{h}_1 \vee \alpha - \hat{g}_1 ) - (\alpha-v_2-\hat{g}_1)^+
\rp^+.\eeqa

\beqa
s_2 &\leq& a_{21} \df \mathrm{min}_{\hat{\A}} \lp 1-\hat{h}_2 - v_2 - (\alpha - v_1-\hat{g}_2)^+ \rp^+ \\
t_2+s_2 &\leq& a_{22} \df \mathrm{min}_{\hat{\A}} \lp 1-\hat{h}_2 - (\alpha - v_1-\hat{g}_2)^+ \rp^+ \\
t_1+s_2 &\leq& a_{23} \df \mathrm{min}_{\hat{\A}} \lp ( 1-\hat{h}_2 - v_2 \vee \alpha - \hat{g}_2 ) - (\alpha - v_1-\hat{g}_2)^+ \rp^+ \\
t_1+t_2+s_2 & \leq& a_{24} \df \mathrm{min}_{\hat{\A}}
\lp(1-\hat{h}_2 \vee \alpha - \hat{g}_2 ) - (\alpha -
v_1-\hat{g}_2)^+ \rp^+.\label{eq:hkrates8} \eeqa

The multiplexing gain achieved by each user is given by \beq r_{i} =
s_{i}+t_{i},\quad i=1,2. \eeq The symmetric multiplexing gains
achievable with the power split $(v_1,v_2)$ is then given by \beq
r(v_1,v_2) = s_{i}+t_{i},\quad i=1,2. \eeq By doing the the
Fourier-Motzkin elimination (as in \cite{CMG08}) it can be shown that,
\begin{align}
r(v_1,v_2) &= \textrm{min} \lbr a_{12}, a_{22}, (a_{11}+a_{24})/2,
(a_{21}+a_{14})/2, (a_{13}+a_{23})/2, \nn \right. \\
&\qquad\qquad \left. (a_{11}+a_{14}+a_{23})/3,
(a_{21}+a_{24}+a_{13})/3 \rbr. \label{eq:rsym}
\end{align}
Note that the non-negativity constraints on the sub-rates need not
be taken into consideration for the Fourier-Motzkin elimination (see
\cite{RPV08},Section 6). The power-split must be chosen so as to
maximize the symmetric rate. Therefore \beq
r_{\mathrm{HK}}(\alpha,d) = \textrm{max}_{(v_1,v_2)} r(v_1,v_2).
\eeq This is a non-convex optimization problem. It can be shown that
when $v_{1}=v_{2}=v$, the fundamental solution is as shown in Table
\ref{tab:HK}. (Simulations suggest that letting $v_{1}\neq v_{2}$
does not increase the symmetric rate any further.)

\section{Worst Case Outer Bound for DMT}\label{app:WCob}

We get an outer bound to the DMT by assuming that the
transmitter knows the channel coefficients $\lp \hat{h}_1,
\hat{g}_1, \hat{h}_2, \hat{g}_2 \rp$. The transmitters can therefore
adopt the power split and the rate split accordingly. We have
already mentioned that the power splits must now be chosen to be
\beq v_1 = \alpha - \hat{g}_2,\qquad v_2 = \alpha - \hat{g}_1. \eeq

Similar to \eqref{eq:hkrates1}-\eqref{eq:hkrates8}, we obtain the
constraint on the sub-rates as \beqa
s_1 &\leq& \bar{a}_{11} \df  \lp 1-\alpha - \hat{h}_1 + \hat{g}_2    \rp^+  \\
t_1+s_1 &\leq& \bar{a}_{12} \df  \lp 1-\hat{h}_1 \rp^+  \\
t_2+s_1 &\leq& \bar{a}_{13} \df  \lp ( 1-\alpha - \hat{h}_1 + \hat{g}_2 \vee \alpha - \hat{g}_1 )  \rp^+ \\
t_2+t_1+s_1 & \leq& \bar{a}_{14} \df  \lp (1-\hat{h}_1 \vee \alpha -
\hat{g}_1 )  \rp^+ \eeqa

\beqa
s_2 &\leq& \bar{a}_{21} \df  \lp 1-\alpha - \hat{h}_2 +\hat{g}_1  \rp^+ \\
t_2+s_2 &\leq& \bar{a}_{22} \df  \lp 1-\hat{h}_2  \rp^+ \\
t_1+s_2 &\leq& \bar{a}_{23} \df  \lp ( 1-\alpha - \hat{h}_2 +\hat{g}_1 \vee \alpha - \hat{g}_2 )  \rp^+  \\
t_1+t_2+s_2 & \leq& \bar{a}_{24} \df  \lp(1-\hat{h}_2 \vee \alpha -
\hat{g}_2 )  \rp^+ \eeqa Similar to \eqref{eq:rsym} the symmetric
rate for a given value of the channel gains is given by,
\begin{align}
r(\hat{h}_1, \hat{g}_1, \hat{h}_2, \hat{g}_2) &= \textrm{min} \lbr
\bar{a}_{12}, \bar{a}_{22}, (\bar{a}_{11}+\bar{a}_{24})/2,
(\bar{a}_{21}+\bar{a}_{14})/2, (\bar{a}_{13}+\bar{a}_{23})/2, \nn \right. \\
&\qquad\qquad \left. (\bar{a}_{11}+\bar{a}_{14}+\bar{a}_{23})/3,
(\bar{a}_{21}+\bar{a}_{24}+\bar{a}_{13})/3 \rbr. \label{eq:rout}
\end{align}
This quantity is now minimized  over the set $\hat{\A}$ to obtain an outer bound
to the DMT:
\beq \hat{r}_{\mathrm{out}}(\alpha,d) =
\textrm{min}_{\hat{\A}}\;r(\hat{h}_1, \hat{g}_1, \hat{h}_2,
\hat{g}_2). \eeq It can be shown that \beq
\hat{r}_{\mathrm{out}}(\alpha,d) = (1-d) \wedge \lp
1-\frac{\alpha}{2}-d \vee \frac{1-d+\alpha}{4} \rp \wedge \lp \alpha
-d \wedge 1-\alpha-d \wedge \frac{1+\alpha-d}{3} \rp. \eeq Observe that
the outer bound obtained here is tighter than the outer bounds given
in \cite{AL08}.

\section{Proof of Theorem \ref{thm:DMTZIFC}} \label{app:DMTZIFC}

We follow an analysis similar to that in Appendix \ref{app:HKscheme} to
characterize the DMT region achieved by the Han-Kobayashi scheme. 
We will derive inner and outer bounds for the general case with $\beta_{1}, \beta_{2}, \alpha_{1}$ and for diversity $d_{1}$ and $d_{2}$, and then compare the bounds for the special case. 
We need to find the g.d.o.f.~region of the compound channel defined by
the no-outage set \beq \hat{\A} = \lbr \hat{h}_1 + \hat{g}_1 \leq d_{1}
\rbr \times \lbr \hat{h}_2 \leq d_{2} \rbr, \label{eq:A} \eeq For
transmitter 1 there is no public message, i.e.~$v_{1}=0$. Its rate
is given by $r_{1} \log \SNR$. For transmitter 2 we use power split
$v_{2}=\alpha_{1}$ with multiplexing gains of the private and public
messages to be $s_{2}\log\SNR$ and $t_{2}\log\SNR$ respectively,
such that \beq r_{2} =s_{2}+t_{2}.\eeq

As earlier, from the decodability constraints at the two receivers, we
have
\begin{align}
r_1 &\leq \mathrm{min}_{\hat{\A}} \lp \beta_{1} - \hat{h}_{1}\rp^+ = \lp \beta_{1} - d_{1} \rp^{+} \\
t_2+r_1 & \leq \mathrm{min}_{\hat{\A}} \lp \beta_{1}-\hat{h}_1 \vee
\alpha_{1} - \hat{g}_1 \rp^+ =
\lp\beta_{1}-d_{1}\vee\alpha_{1}-d_{1}\vee\frac{\beta_{1}-d_{1}+\alpha_{1}}{2}\rp^{+}.
\end{align}
\begin{align}
s_2 &\leq \mathrm{min}_{\hat{\A}} \lp \beta_{2}-\hat{h}_2 - \alpha_{1} \rp^+ =(\beta_{2}-\alpha_{1}-d_{2})^{+} \\
t_2+s_2 &\leq \mathrm{min}_{\hat{\A}} \lp \beta_{2}-\hat{h}_2 \rp^+ =
(\beta_{2}-d_{2})^{+}.
\end{align}

By Fourier-Motzkin elimination, the achievable g.d.o.f.~with the
given Han-Kobayashi scheme is given by
\begin{align}
r_{1}&\leq (\beta_{1}-d_{1})^{+}, \nn\\
r_{2} &\leq (\beta_{2}-d_{2})^{+}, \nn\\
r_{1}+r_{2} &\leq
\lp\beta_{1}-d_{1}\vee\alpha_{1}-d_{1}\vee\frac{\beta_{1}-d_{1}+\alpha_{1}}{2}\rp^{+}
+(\beta_{2}-\alpha_{1}-d_{2})^{+}.
\end{align}

We next consider the worst case outer bound.
 When the fading level is $\hat{a}=(\hat{h}_{1},\hat{h}_{2},\hat{g}_{1}) \in
\hat{\A}$, the g.d.o.f.~region for the Z-interference channel
$\C{R}(\hat{a})$ is given by
\begin{align}
r_{1}&\leq (\beta_{1} - \hat{h}_{1})^{+}, \nn\\
r_{2} &\leq (\beta_{2}-\hat{h}_{2})^{+}, \nn\\
r_{1}+r_{2} &\leq \lp \beta_{1}-\hat{h}_1 \vee \alpha_{1} -
\hat{g}_1 \rp^+ + (\beta_{2}-\alpha_{1}-\hat{h}_{2}+\hat{g}_{1})^{+}.
\end{align}
The worst case outer bound is then given by
\begin{align}
\C{R}_{o}(d_{1},d_{2})  &=  \bigcap_{\hat{\A}}  \C{R}(\hat{a}) \nn\\
&=\lbr\;r_{1} \leq \right.\textrm{min}_{\hat{\A}}\;(\beta_{1} - \hat{h}_{1})^{+} = \lp \beta_{1}-d_{1} \rp ^{+} , \nn\\
&\qquad r_{2} \leq\textrm{min}_{\hat{\A}}\; (\beta_{2}-\hat{h}_{2})^{+} = \lp \beta_{1}-d_{1} \rp ^{+}, \nn\\
&\qquad r_{1}+r_{2} \leq \left. \textrm{min}_{\hat{\A}}\; \lp \beta_{1}-\hat{h}_1 \vee \alpha_{1} - \hat{g}_1 \rp^+ + (\beta_{2}-\alpha_{1}-\hat{h}_{2})^{+} = K \rbr,
\end{align}
where 
\beq 
K = \lp \beta_{2}-d_{2} \vee \frac{\beta_{1}-\alpha_{1}-d_{1}}{2} \vee \beta_{1}-d_{1} \vee \beta_{1}+\beta_{2}-\alpha_{1}-d_{1}-d_{2} \vee \alpha_{1}-d_{1} \rp ^{+}.
\eeq
It is easy to verify that when $d_{1}=d_{2} =d$ and $\beta_{1}=\beta_{2}=1$, the worst case outer bound meets the inner bound 
\hfill$\Box$

\section*{Acknowledgments}
The authors would like to thank Vaneet Aggarwal for pointing out errors in an earlier version of this paper.

\end{document}